\documentclass[12pt]{iopart}
\expandafter\let\csname equation*\endcsname=\relax 
\expandafter\let\csname endequation*\endcsname=\relax 
\usepackage{amsmath}

\usepackage{graphicx}
\usepackage{color}
\newcommand{\revised}[1]{{#1}}
\newcommand{\revisedd}[1]{{#1}}

\newcommand{\ud}{\, \mathrm{d}}

\begin{document}
\title{Patterned Random Matrices: deviations from universality}
	\author{Md. Sabir Ali$ ^1 $,
	Shashi C. L. Srivastava$ ^{1,2} $}
\address{$^{1}$Variable Energy Cyclotron Centre, Kolkata 700064, India.}
\address{$^2$Homi Bhabha National Institute, Training School Complex,
Anushaktinagar, Mumbai - 400094, India}

 \ead{shashi@vecc.gov.in}

\begin{abstract}
\revised{We investigate the level spacing distribution for three 
ensembles of real symmetric matrices having additional structural 
constraint to reduce the number of independent entries to only  $ 
(n+1)/2 $ in contrast to the $ n(n+1)/2 $ for a  real 
symmetric matrix of size $ n \times n $. We derive all the results 
analytically exactly for the $ 3\times 3 $ matrices  and show that 
spacing distribution display a range of behaviour based on the 
structural constraint. The spacing distribution of the ensemble of 
reverse circulant matrices with additional zeros is found to fall 
slower than exponential for larger spacing while that of symmetric 
circulant matrices has poisson spacings. The palindromic symmetric 
toeplitz matrices on the other hand show level repulsion but the 
distribution is significantly different from Wigner. The behaviour  of 
spacings for all the three ensembles clearly show 
the departure from universal result of Wigner distribution for real 
symmetric matrices. The deviation from universality continues in large $ 
n $ cases as well, which we study numerically.}
\end{abstract}

\vspace{2pc}

	\maketitle
\section{Introduction}
Wigner conceptualized the random matrix theory (RMT) as a theory to 
explain fluctuation properties of the system, based solely on the 
symmetry 
properties of the Hamiltonian \cite{Wig1957a}. Since then RMT has been 
successfully applied in studies across the fields of Nuclear physics, 
condensed matter physics, mathematics\cite{GuhMueWei1998,RMTScholarpedia}. 
The 
Bohigas-Giannoni-Schmit (BGS )
conjectured that fluctuation properties of spectrum of classically chaotic 
systems 
are 
described by the RMT ensembles dictated by the symmetry of the 
Hamiltonian\cite{BohGiaSch1984} while Poisson distributed level spacings 
are 
characteristics of a generic integrable system \cite{BerTab1977}. 
Dyson's three fold 
classification of the RMT ensembles based on the invariance of probability 
measure under the symmetry group have since then been 
generalized to 10 fold symmetry based classification 
\cite{Dys1962b,AltZir1997}. The universality of random matrix theory is 
attributed to 
such 
symmetry based approach. The universality is such a powerful tool that any 
deviation from it hints to some specific feature of the physical system.

Patterned matrices such as Toeplitz, Hankel, cyclic (circulant) 
matrices have attracted a lot of attention in mathematics and statistics 
community. The book \cite{Bos2018} summarizes a lot of available 
results. The quantity of interest which has been looked most often are 
limiting spectral distributions (in physics literature commonly used 
term is eigenvalue density) and various moments \cite{BosSahSen2021, 
BosSahSen2022, AdhSah2017, BlaBorBos2021}. Circulant matrices are  
unarguably the most 
studied one in this family. This class of matrices have also been 
studied from the perspective of pseudo-Hermiticity, as well as 
constrained systems where constraints put on the matrices over and above 
the constraints put by symmetry requirements of the ensemble 
\cite{JaiSri2008, JaiSri2009, SriJai2012, ShuSad2015, SadShu2015}. A 
number of physical scenarios where such a constraint can occur are 
discussed in \cite{ShuSad2015}.

In this paper we study three ensembles of real symmetric matrices, each 
having only $ \frac{n+1}{2} $ number of independent elements in contrast 
to $ \frac{n(n+1)}{2} $ independent elements of a real symmetric matrix 
constrained by orthogonal symmetry. We study the representative $ 
3\times 3 $ matrices from each ensemble \revised{analytically exactly} 
and indicate the results for 
larger dimensionality wherever possible. \revised{We compliment these 
studies numerically for not only $3\times 3  $ matrices but for general 
$ n \times n $ dimensional matrices as well.} The matrix elements 
are chosen to be independently Gaussian distributed subjected to the 
constraint imposed by the symmetry and structure of the matrix. Our 
quantity of interest here is the consecutive spacing distribution (also 
known as nearest neighbour spacing distribution) of the 
two independent eigenvalues \revised{for $ 3\times 3 $ case which we 
appropriately generalize to nearest neighbour spacing distribution of 
symmetry reduced spectrum in case of $ n\times n $ dimensional 
matrices.} 
These 
ensembles themselves find place in studying the systems for which 
Hamiltonian is translationally invariant and has coupling matrices as 
cyclic matrix \cite{GluEisFar2019}, a particle doing the Markovian walk 
on a circle 
\cite{ManSriJai2011}, or the problem of studying Ohm's law on a disk 
\cite{Dem2017}.
We define and study the variation of ensemble of reverse cyclic, 
symmetric cyclic and palindromic symmetric Toeplitz matrices in Section 
2, 3 and 4. The details of spacing distribution calculations are 
presented for $ 3\times 3 $ matrices and contrasting results are shown 
for the three ensembles despite them being the symmetric matrix with 
exactly two independent elements. \revised{The numerical results for $ n 
\times n $ dimensional matrices of the respective class are also 
presented and similarity/deviation from their $ 3\times 3 $ analogue is 
discussed.}

\section{Random reverse cyclic matrices with additional zeros}
	Consider an ensemble of reverse cyclic (RC) matrices, drawn from a 
	Wishart distribution,
	\begin{equation}\label{eq:ph-w}
		P(H)\sim \exp\left(-A \mbox{Tr}(H^\dag H\right)),
	\end{equation}
	with $ A $ as normalization constant.
	Let us start with the simplest case, namely an ensemble of $3\times 
	3$ reverse cyclic matrices with one entry $ c=0 $ \footnote{This 
	ensemble of size $ n \times n $ matrices can be constructed by 
choosing first $ (n+1)/2 $ from independent Gaussian distributions while 
rest $ (n-1)/2 $ entries are set to zero.} 
	\begin{equation}
		H = \begin{pmatrix} a & b & c\\
			b & c & a\\
			c & a & b \end{pmatrix}.
	\end{equation}
	Using (\ref{eq:ph-w}), the joint probability distribution function 
	(JPDF) in matrix space  
	will be given by,
	\begin{equation}\label{eq:ph3-rc}
		P(a, b, c) = \left(\frac{3 A}{\pi}\right) \exp [-3 A 
		\left(a^2+b^2+c^2\right)]\delta(c).
	\end{equation}
	
\begin{figure}[h]
\centering
\includegraphics[width=0.48\textwidth]{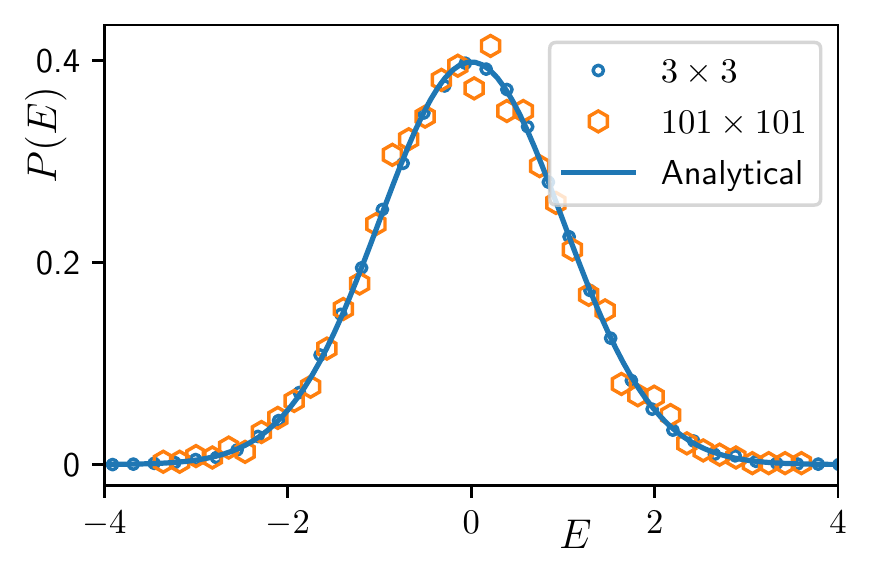}
\includegraphics[width=0.48\textwidth]{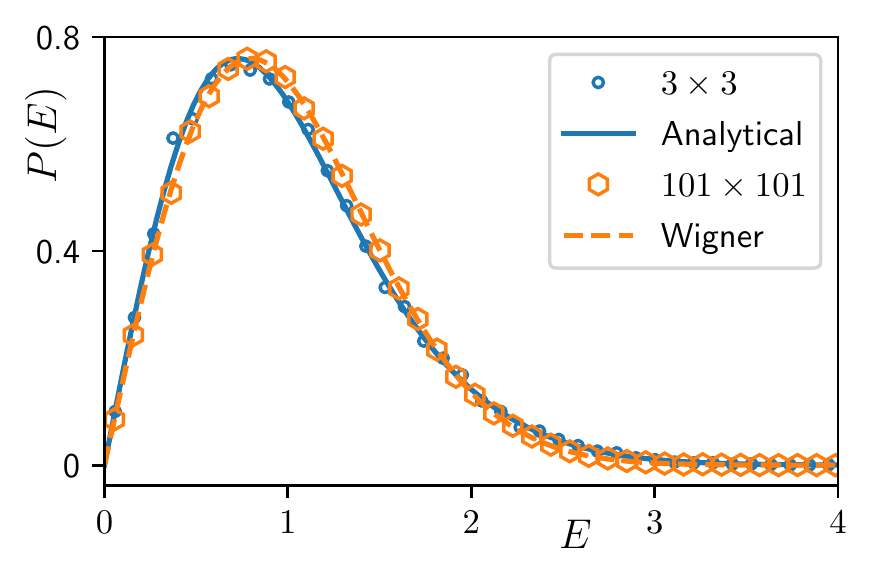}
\caption{(Left) For $ 3\times 3 $ reverse-circulant matrices with $ c=0 
	$, numerical \revisedd{density distribution} of the trivial eigenvalue 
	with 
	sample 
	size 50000 \revised{ is shown by blue open circle}. The result agrees 
	well with the 
	analytical prediction obtained in Eq. \ref{eq:density_triv_eigv} 
	\revisedd{(represented by the solid blue line)}. 
	\revised{\revisedd{Orange} 
		hexagons represent the \revisedd{numerical} density of trivial 
		eigenvalue for $ 
		101\times 101 
		$ matrices of this class with sample size 4000.}  (Right) For $ 
		3\times 
	3 
	$ reverse-circulant matrices with $ c=0 $ \revisedd{(represented by 
	blue open circles)} \revised{as well as for $ 101 
		\times 101 $ matrices of this class} \revisedd{(represented by 
		orange hexagons)}, 
	numerical density \revisedd{distribution }of the non-trivial positive 
	eigenvalue 
	with sample size 50000 \revised{and 4000 respectively}. The result 
	\revised{for $ 3\times 3 $} agrees well with the analytical 
	prediction \revisedd{(represented by solid blue line)} obtained in Eq. 
	\ref{eq:density_nontrivial_eigv} 
	\revised{while density for $ 101 \times 101 $ goes 
		to Wigner distribution} \revisedd{(represented by orange dashed 
		line).}}
\label{fig:eigv_density-rc} 
\end{figure}

	Karner \textit{et al.} \cite{KarSchUeb2003} have shown that the 
	eigen-decomposition for an odd-dimensional reverse cyclic  matrix  is 
	given by
	\begin{eqnarray}\label{eq:decomp}
		H &=& F^\dagger \begin{pmatrix}1 & 0\\ 0 & R\end{pmatrix} 
		\Lambda \begin{pmatrix}1 & 0\\ 0 & R^\dagger \end{pmatrix} 
		F\\ \nonumber
		\Lambda &=& (E_1, |E_2|,\ldots, 
		|E_{(n-1)/2}|,-|E_{(n-1)/2}|,\ldots, -|E_2|),
	\end{eqnarray} 
	where $ F $ is Fourier matrix.
  For the $3 \times 3$ case, the explicit form of $R$ is 
  \begin{equation}
  	R = \begin{pmatrix}
  		 \frac{1}{\sqrt 2} \exp(-i \theta/2) & \frac{i}{\sqrt 2} \exp(-i 
  		\theta/2)\\
  		 \frac{1}{\sqrt 2} \exp(i \theta/2) & -\frac{i}{\sqrt 2} \exp(i 
  		\theta/2) \end{pmatrix}.
  \end{equation}
  It takes a simple algebra then to show that
  \begin{eqnarray}\label{eq:abc_eig}
  	\nonumber
  	a&=&\frac{1}{3} (E_1+2 |E_2|\cos \theta) \\
  	b&=&\frac{1}{3} \left( E_1- |E_2| \left(\cos \theta +\sqrt 3 \sin 
  	\theta \right)\right) \\ \nonumber
  	c&=&\frac{1}{3} \left(E_1- |E_2|\left(\cos \theta -\sqrt 3 \sin \theta 
  	\right)\right).
  \end{eqnarray}
Using (\ref{eq:abc_eig}) in (\ref{eq:ph3-rc}), we can find the JPDF for 
eigenvalues and an independent parameter $\theta$ coming from the 
eigenvector. Note that in $H$, the independent parameters are three in 
number, namely $a,b~\mbox{and}~c$; while in the eigen-decomposition, 
we have $E_1, E_2, \theta$. The Jacobian for the transformation 
(\ref{eq:abc_eig}) is given by $\frac{2 \left|E_2\right|}{3 \sqrt 3}$.
The JPDF for eigenvalues is 
\begin{equation}
	\begin{aligned}
		P(E_1,|E_2|,\theta) &= \frac{2 |E_2|}{3 \sqrt 3} \left(\frac{3 
	A}{\pi}\right) \exp [-A \left(E_1^2 + 2 E_2^2\right)] 
	\delta\left[\frac{1}{3} \left(E_1- 2|E_2|\sin \left(\frac{\pi}{6} - 
	\theta \right)\right)\right]\\ 
	\nonumber
	&~ ~~\mbox{where}~ E_1 \in (-\infty,\infty), |E_2|\in [0,\infty), ~ 
	\theta \in [0,2\pi).
	\end{aligned}
\end{equation}
Notice that the domain of $|E_2|$ is $[0,\infty )$, and that the function 
on the right hand side is an even function of $E_2$. Using the following 
identity for $ \delta- $ function, 
\[ \delta(f(\theta))  = \sum_n\frac{\delta (\theta - 
\theta_n)}{f'(\theta_n)},\] 
and the fact that between $ [0, 2\pi) $, we will have two roots, we can 
rewrite the JPDF after an integration over $\theta$ in the 
following form,
\begin{equation}\label{eq:pe1e2_int}
	P(E_1,E_2) = \frac{2|E_2|}{3 \sqrt 3} \left(\frac{3 
	A}{\pi}\right) \exp(-A \left(E_1^2 + 2 E_2^2\right)) 
	\int_{0}^{2\pi} \sum_n \frac{\delta(\theta - 
	\theta_n)}{|f'(\theta_n)|} 
	d\theta
\end{equation}
The zeros of the delta function argument occur at,
\[ \theta_1 = \frac{1}{6}\left[\pi -6\sin ^{-1}\left(\frac{E_1}{2 \left| 
E_2\right| }\right)\right], \quad \theta_2 = \frac{1}{6}\left[\pi 
-6\left(\pi - \sin ^{-1}\left(\frac{E_1}{2 \left| 
E_2\right| }\right)\right)\right] \]
and the derivative of the function evaluated at these values are,
\[ \left|f'(\theta_1)\right| = \frac{1}{3} \sqrt{4 \left| 
E_2\right|^2-E_1^2}, \quad  \left|f'(\theta_2)\right| = \frac{1}{3} 
\sqrt{4 \left| 
E_2\right|^2-E_1^2}.\]
Using these, (\ref{eq:pe1e2_int}) becomes,
\begin{equation}\label{eq:jpdf-e1e2-rc}
	P(E_1, E_2) = \begin{cases}
		\frac{2|E_2|}{3 \sqrt 3} \left(\frac{3 
			A}{\pi}\right) \exp(-A \left(E_1^2 + 2 E_2^2\right))  
		\frac{6}{\sqrt{4 \left| 
				E_2\right|^2-E_1^2}} & \left|\frac{E_1}{2 
			\left| E_2\right| }\right| \leq 1 .\\
		0 & \text{ otherwise }.
	\end{cases} 
\end{equation}
The density distribution of $ E_1 $ \revised{which we will refer as 
trivial eigenvalue being the sum of all the elements in the first row,} 
can be found out to be,
\begin{equation}\label{eq:density_triv_eigv}
	\begin{aligned}
		P(E_1) &= \int P(E_1, E_2) dE_2\\
		&= \int_{\frac{|E_1|}{2}} ^\infty \frac{4|E_2|}{\sqrt{ 3} \sqrt{4 
		\left| E_2\right|^2-E_1^2}} \left(\frac{3 A}{\pi}\right)
		\exp(-A \left(E_1^2 + 2 E_2^2\right)) dE_2\\
		&=\sqrt{\frac{3 A}{2\pi }} e^{-\frac{3}{2} A E_1^2},
	\end{aligned}
\end{equation}
while density distribution of $ E_2 $ \revised{which we will refer as 
non-trivial positive eigenvalue} is,
\begin{equation}\label{eq:density_nontrivial_eigv}
	\begin{aligned}
		P(E_2) &= \int P(E_1, E_2) dE_1\\
		&= \int_{-2E_2} ^{2E_2} \frac{4|E_2|}{\sqrt{ 3} \sqrt{4 
				\left| E_2\right|^2-E_1^2}} \left(\frac{3 A}{\pi}\right)
		\exp(-A \left(E_1^2 + 2 E_2^2\right)) dE_1\\
		&=4 \sqrt{3}A E_2 e^{-4 A E_2^2} I_0\left(2 A E_2^2\right).
	\end{aligned}
\end{equation}
The free parameter $ A $ is fixed by setting the mean energy $ \langle 
E_2 \rangle$ as 1. Such a distribution is compared 
with numerics in Fig.\ref{fig:eigv_density-rc}. 
\revised{The density of trivial eigenvalue for the $ n$-dimensional 
matrices remains Gaussian while density of non-trivial (positive) 
eigenvalues deviates from analytical result obtained in Eq. 
\ref{eq:density_nontrivial_eigv} and transits to Wigner distribution. 
This is  
in agreement with the result of random reverse circulant matrices with $ 
n $-independent elements \cite{SriJai2012}. These results are clearly 
borne out from numerical data presented in Fig. 
\ref{fig:eigv_density-rc}. The fluctuation seen for trivial eigenvalue 
density in the case of $ 101 \times 101 $ is due to smaller sample size 
of $ 4000 $ considered here as for each sample we obtain only one 
trivial eigenvalue and fifty non-trivial positive eigenvalue. }

\subsection{Spacing distribution}
\begin{figure}[ht]
	\centering
	\includegraphics[width=0.48\textwidth]{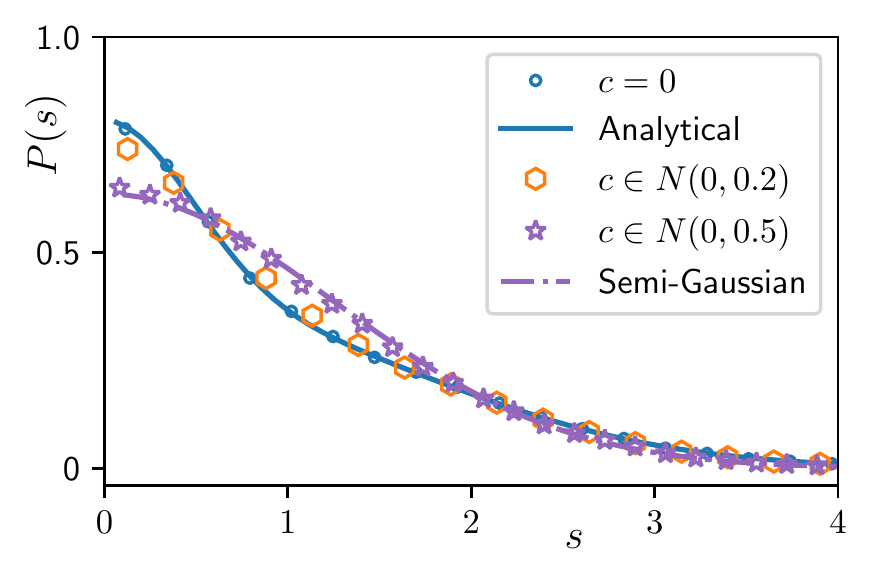}
	\includegraphics[width=0.48\textwidth]{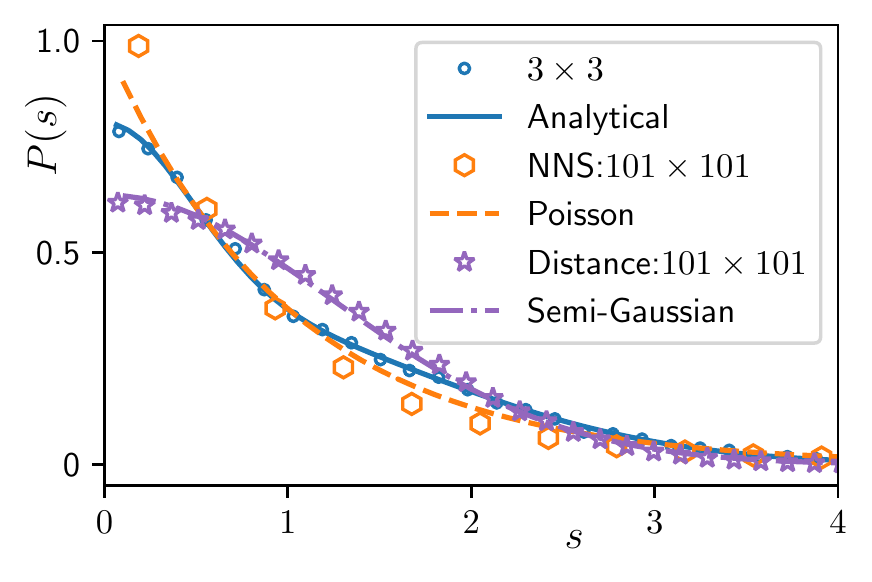}
		\caption{\revised{(Left) Plot of transition in $P(s)$ for $ 
		3\times 
			3 $ reverse-circulant matrices as we ``soften'' the constraint 
			on $ 
			c $ from 0 to a Gaussian distributed random number with mean 0 
			and 
			finite variance. \revisedd{Blue open circles represent the 
			behaviour for zero variance \textit{i.e.} c=0. Orange 
			hexagons and purple stars represent the behaviour for 
			variances $\sigma=0.2$ and $\sigma=0.5$ respectively. In 
			both the panels, solid blue line represents the spacing 
			distribution calculated analytically (\ref{eq:ps-rc}) and 
			the purple dash-dot 
			line represents semi-Gaussian curve (\ref{eq:semi-gauss})}. 
			(Right) Plot 
			of probability distribution function (PDF) of spacing between 
			trivial eigenvalue and non-trivial positive 
			eigenvalue for $ 3\times 3 $ case \revisedd{(represented by 
			blue open circles)} along with PDF of distance between 
			trivial eigenvalue to each of the positive non-trivial 
			eigenvalue 
			for $ 101 \times 101 $ size matrices \revisedd{(represented by 
			purple stars)}. The nearest neighbour spacing  
			distribution of positive eigenvalues is also plotted for $ 
			101\times 
			101$ which tends towards Poisson 
			distribution \revisedd{(represented by orange hexagons)}. 
			\revisedd{Orange dashed line represents a Poisson 
			distribution. In all the plots, marker symbols are used for 
			numerically obtained data, while lines are respective 
			analytical results.}  
			}}\label{fig:3by3spacing_plot}
\end{figure}

Let's define the spacing, $ s $ between trivial eigenvalue $ E_1 $ and 
the positive non-trivial eigenvalue $ E_2 $ as $ s= \left|E_1-E_2\right| 
$. Then the distribution of $ s $ is defined using JPDF 
(\ref{eq:jpdf-e1e2-rc}) as,
	\begin{equation}\label{eq:spgn}
		P(s)=\int P(E_1, E_2) \delta(s-|E_1-E_2|) \ud E_1 \ud E_2
	\end{equation}
We can rewrite equation (\ref{eq:spgn}) in terms of eigenvalues 
explicitly using equation (\ref{eq:jpdf-e1e2-rc}) and after applying 
proper limits one gets,

\begin{equation}\label{eq:sp1}
	\begin{aligned}
		P(s) &=\frac{4\sqrt{3} A}{\pi} \int_{0}^{\infty}E_2 \exp 
		[-2Ae_2^2]  \int_{-2E_2}^{2E_2} \frac{\exp 
		[-Ae_1^2]}{\sqrt{4E_2^2-E_1^2}} \delta(s-|E_1-E_2|) \ud E_1\ud 
		E_2\\
		&= k~ \int_{0}^{\infty}E_2 \exp [-2Ae_2^2] I(E_2) \ud E_2
	\end{aligned}
\end{equation}
where $k=\frac{4\sqrt{3} A}{\pi}$ and $I(E_2)$ is defined as,
\begin{equation}\label{eq:integ2}
	I(E_2) = \int_{-2E_2}^{2E_2} \frac{\exp 
	[-AE_1^2]}{\sqrt{4E_2^2-E_1^2}} 
	\delta(s-|E_1-E_2|) \ud E_1. 
\end{equation} 
Taking care of the domain of JPDF of eigenvalues, equation 
(\ref{eq:integ2}) can be broken into two parts with respective 
limits as shown below,
\begin{equation}\label{eq:i1i2}
	\begin{aligned}
		I(E_2) &=\int_{-2E_2}^{E_2} \frac{\exp 
		[-AE_1^2]}{\sqrt{4E_2^2-E_1^2}} 
		\delta(s-(E_2-E_1)) \ud E_1 + \int_{E_2}^{2E_2} \frac{\exp 
		[-AE_1^2]}{\sqrt{4E_2^2-E_1^2}} \delta(s-(E_1-E_2)) \ud E_1\\
		&= I_1+I_2 
	\end{aligned}
\end{equation}
$I_1$ and $I_2$ can be evaluated separately under certain conditions 
which are stated below,
\begin{equation}\label{eq:i}
	\begin{aligned}
I_1 &= \begin{cases}\frac{\exp [-A 
		(e_2-s)^2]}{\sqrt{4e_2^2-(e_2-s)^2}}& \text{for }e_2 \geq 
		\frac{s}{3} \text{ and }s\geq 0, \\
			0 &\text{otherwise}\end{cases} \\
I_2 &= \begin{cases}\frac{\exp [-A 
		(e_2+s)^2]}{\sqrt{4e_2^2-(e_2+s)^2}}& \text{for }e_2 \geq s 
		\text{ 
		and }s\geq 0, \\
			0 &\text{otherwise}\end{cases}.
	\end{aligned}
\end{equation}
Utilizing (\ref{eq:i}), the spacing distribution is,
\begin{equation}\label{eq:ps-rc}
	\begin{aligned}
		P(s) =\frac{4A}{\pi}  \exp \left[-\frac{4A}{6} s^2 \right] 
		\left[\int_0^{\infty} \frac{\left(x+\frac{s}{3}\right) \exp 
		\left[-3Ax^2\right]}{\sqrt{x\left(x+\frac{4s}{3}\right)}} \ud x 
		\right.
		+\left.\int_{\frac{4s}{3}}^{\infty} 
		\frac{\left(x-\frac{s}{3}\right) \exp 
		\left[-3Ax^2\right]}{\sqrt{x\left(x-\frac{4s}{3}\right)}} \ud 
		x\right]
	\end{aligned}
\end{equation}
For $s=0$ one can show from eq. (\ref{eq:ps-rc}) that, 
\begin{equation}
	P(s\to 0)=\frac{4}{\sqrt{3}}\sqrt{\frac{A}{\pi}}.
\end{equation}
The probability distribution of spacing deviates from Poisson and Wigner 
distribution expected from orthogonal symmetry. Such deviations from 
universal results are characteristic feature of patterned random 
matrices. The behaviour of $ P(s) $ obtained analytically in 
(\ref{eq:ps-rc}) is compared against the numerical calculation in 
Fig. \ref{fig:3by3spacing_plot}. The agreement between the two is very 
good.

\revised{To separate the role played by ``hard'' constraint of putting $ 
c=0 $ from the symmetry, we numerically studied the distribution of 
spacing between trivial eigenvalue and non-trivial positive eigenvalue 
by choosing $ c $ as a normal distributed random number with mean 0 and 
finite variance. Note that variance approaching to zero is equivalent 
to  $ c \to 0 $. The distribution goes from Eq. \ref{eq:ps-rc} 
to semi-Gaussian distribution as we increase the variance . This 
transition is clearly visible in Fig. \ref{fig:3by3spacing_plot} (Left 
Panel). For comparable variances of all the three 
elements, the matrix is in random reverse circulant matrix class with 
all the three independent elements and the distribution comes out to be 
semi-Gaussian expectedly as derived in 
\cite{SriJai2012} for random reverse circulant matrices. Moreover, we 
will see that constraint of putting the two elements namely $ b $ and $ 
c $ equal also yields the semi-Gaussian as obtained in Eq. 
\ref{eq:semi-gauss}. This clearly shows that hard constraint of putting 
$ c=0 $ fundamentally change the behaviour of spacings. Such behaviour 
when hard constraints affect the 
spacing distribution has been studied in the case of random banded 
matrices where level repulsion develops logarithmic singularity due to 
constraint \cite{MolSok1989}.}

\revised{We extend our studies to large matrix sizes numerically. The 
spectrum of these constrained random reverse circulant matrices has 
always one eigenvalue as the sum of the first row elements, and rest 
eigenvalues come in $ \pm $ pair. As we are interested in symmetry 
reduced spectrum in spacing distribution, we leave out all the negative 
eigenvalues from the $ \pm $ pair. We study two quantities (i) distance 
between trivial eigenvalue to each of the positive non-trivial 
eigenvalue and (ii) the nearest neighbour spacing (NNS) between 
non-trivial positive eigenvalues. It turns out that in large $ n $ 
(dimensionality of matrix) limit the probability distribution of 
distance between trivial to positive non-trivial eigenvalue becomes 
semi-Gaussian, a result obtained for the case of unconstrained random 
reverse circulant matrices. The NNS on the other hand tend to become 
Poisson distribution as seen in Fig. \ref{fig:3by3spacing_plot} (Right 
Panel). Clearly 
reducing the independent elements from $ n $ to $ (n+1)/2 $ is not 
enough in the large $ n $ limit to affect the spacing distribution to 
deviate from random reverse circulant matrices. Having said that, the 
spacing distribution is far from Wigner distribution expected from real 
symmetric matrices.}

\section{Symmetric circulant matrices}
\begin{figure}[ht]
	\centering
	\includegraphics[width=0.75\textwidth]{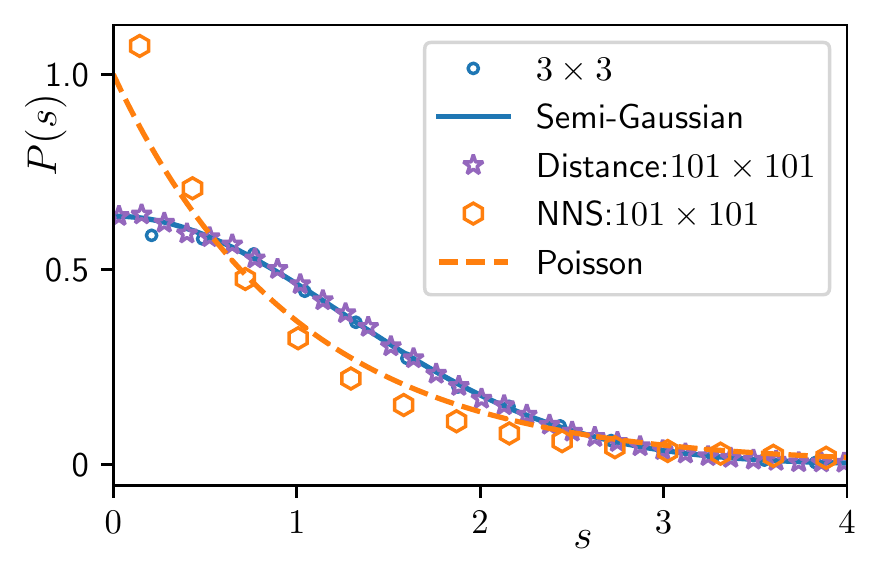}
	\caption{\revised{ Plot of probability distribution function 
			(PDF) of spacing between even parity eigenvalues for $ 3\times 
			3 $ 
			case \revisedd{(represented by blue open circles)} along with 
			PDF of distance between 
			trivial eigenvalue to each of non-trivial even parity 
			eigenvalues 
			for $ 101 \times 101 $ size matrices \revisedd{(represented by 
			purple stars)}. The nearest neighbour spacing  
			distribution of even parity symmetry eigenvalues is also 
			plotted for 
			$ 101\times 101 $ which tends towards Poisson 
			distribution \revisedd{(represented by orange hexagons)}. 
			\revisedd{Solid blue line represents a semi-Gaussian function 
			(\ref{eq:semi-gauss}) 
			and the orange dashed line represents a Poisson 
			distribution. Different marker symbols are used for 
			numerically obtained data, while lines are used for respective 
			analytical/functional form.}}}\label{fig:spacing_plot_sym_circ}
\end{figure}

Symmetric circulant matrices naturally occur as a coupling matrix 
whenever Hamiltonian of the system is translationaly  invariant along 
with the fact that it is hermitian \cite{GluEisFar2019}. In a classical 
setting of studying Ohm's law on a disk, the Ohm's matrix relating the 
vectors of voltages and currents is symmetric circulant\cite{Dem2017}. 
Consider an ensemble of such symmetric circulant matrices, drawn from a 
Wishart distribution. 
Let us start again with the simplest case, namely an ensemble of 
$3\times 3$ cyclic (circulant) matrices with additional constraint $ b=c 
$, 
	\begin{equation}
		H = \begin{pmatrix} a & b & c\\
			c & a & b\\
			b & c & a \end{pmatrix}.
	\end{equation}
The $ n\times n $ generalization of the symmetric matrix is obtained 
easily by setting the first row element of a circulant matrix such that 
$ a_{n-i} = a_i, i = 1,\dots (n-1) $. Using (\ref{eq:ph-w}), the JPDF in 
matrix space will be given by,
	\begin{equation}\label{eq:ph3}
		P(a, b, c) = \left(\frac{3\sqrt{2} A}{\pi}\right) \exp [-3 A 
		\left(a^2+b^2+c^2\right)]\delta(b-c).
	\end{equation}
	The eigenvalues of this symmetric circulant matrix is given by,
	\[ \begin{bmatrix}
		E_1\\E_2
	\end{bmatrix} = \begin{bmatrix}
	1 & 2\\1 & 2\cos \frac{2\pi}{3}
\end{bmatrix} \begin{bmatrix}
a\\b
\end{bmatrix}. \]
Using this relation along with the fact that eigenfunctions are columns 
of discrete Fourier matrix, the joint probability distribution of 
eigenvalues can straightforwardly be written as,
\begin{equation}
	P(E_1, E_2) = \frac{\sqrt{2}A}{\pi} \exp(-A(E_1^2 + 2E_2^2)).
\end{equation}
Note that number of independent eigenvalues are only two, the third is 
additive inverse equal to $ E_2 $. \revised{Also $ H $ commutes with a 
parity 
like operator $ \eta $ defined as, 
\[ \eta = \begin{bmatrix}
1 & 0 & 0\\
0 & 0 & 1\\
0 & 1 & 0
\end{bmatrix}  \text{ with } \quad \eta^2 = \begin{bmatrix}
1 & 0 & 0\\
0 & 1 & 0\\
0 & 0 & 1
\end{bmatrix}. \]
This pattern generalizes for $ n\times n $ matrices as well. Due to this 
fact, the spectrum of the matrix separates into two, one with even 
parity while 
the other with odd parity. The two eigenvalues considered here 
correspond to even parity sector. This information is important for the 
calculation of spacing distribution for which we need symmetry reduced 
spectrum.} 
For spacing distribution, we consider only 
non-trivial eigenvalues. The ordering of the eigenvalues before the 
calculation of nearest neighbour distribution is simple in this case 
namely $ E_2 = E_1 \pm s $.  The spacing distribution then is defined as,
\begin{equation}
P(s) = \int p(E_1, E_1+s) dE_1 + \int p(E_1, E_1-s) dE_1
\end{equation}
It is straightforward to evaluate the two Gaussian integration, which 
yields the spacing distribution as,
\begin{equation}\label{eq:gaussian_raw}
P(s) = 2 \sqrt{\frac{2}{3 \pi }} \sqrt{A} e^{-\frac{1}{3} \left(2 A 
s^2\right)}.
\end{equation}
After scaling the spacing in such a way that mean spacing becomes one,  
distribution of scaled spacing is,
\begin{equation}\label{eq:semi-gauss}
P(\tilde{s}) = \frac{2}{\pi } e^{-\frac{\tilde{s}^2}{\pi }}.
\end{equation}
This is the Poisson equivalent for two eigenvalue case for spacing 
distribution \cite{BerShu2009}.  \revised{The $ N\times N $ 
generalization of this spacing distribution is expected to be Poisson. 
Like in constrained reverse circulant matrices, we plot the probability 
distribution of distance between trivial eigenvalue and rest of the even 
parity eigenvalue as well as the nearest neighbour distribution for even 
parity spectrum of $ 101 \times 101 $ size matrices from this ensemble 
in Fig. \ref{fig:spacing_plot_sym_circ}. The distance distribution 
remains semi-Gaussian like $ 3\times 3 $ case, while NNS approaches to 
Poisson distribution.}

A contrasting point to note here is that spacing distribution comes out 
to be very different despite the two ensembles containing real 
symmetric matrices and exactly the same number of independent matrix 
elements \revised{for $3 \times 3  $ matrix, but differences does not 
survive the large $ n $ limit.}

\section{Palindromic symmetric Toeplitz (PST) matrices}
The third ensemble is of palindromic symmetric Toeplitz matrices which are 
studied 
again mainly in mathematics literature for the limiting spectral 
distributions of eigenvalues \cite{BosChaGan2003, BryDemTie2006, 
MasMilSin2007, BlaBorBos2021}. The spacing distribution has been 
conjectured to be Poisson in \cite{MasMilSin2007}. Let's again start 
with the simplest case of $ 3 \times 3 $ matrices.
	\begin{equation}\label{eq:3by3pst}
		H = \begin{pmatrix} a & b & a\\
			b & a & b\\
			a & b & a \end{pmatrix}.
	\end{equation}
The JPDF of the matrix elements when matrices are taken from Wishart 
distribution are written as, 
\begin{equation}\label{eq:explicit_jpdf_of_entries}
	\begin{aligned}
		P(a,b) &=\frac{2\sqrt{  5} A}{\pi} \exp[-A(5a^2+4b^2)].
	\end{aligned}
\end{equation}
The eigenvalues ($E_1$, 
$E_2$ and $E_3$) of $3\times3$ PST matrices are given by,
\begin{equation}\label{eq:eigenvalues}
	\begin{aligned}
		E_1 &=0\\
		E_2 &=\frac{1}{2}(3a+\sqrt{a^2+8b^2})\\
		E_3 &=\frac{1}{2}(3a-\sqrt{a^2+8b^2})
	\end{aligned}
\end{equation}
Finally, using equations (\ref{eq:explicit_jpdf_of_entries}), 
(\ref{eq:eigenvalues}) and computing the jacobian we arrive at,
\begin{equation}\label{eq:explicit_jpdf_of_eigenvalues}
	\begin{aligned}
		P(E_1=0, E_2, E_3) &=\frac{\sqrt{5}A}{2 
		\pi}\frac{\lvert 
		E_3-E_2\rvert}{\sqrt{(E_3-E_2)^2-\frac{1}{2}E_2E_3}}
		 \exp[-A(E_2^2+E_3^2)]
	\end{aligned}
\end{equation}

\begin{figure}[ht]
	\centering
	\includegraphics{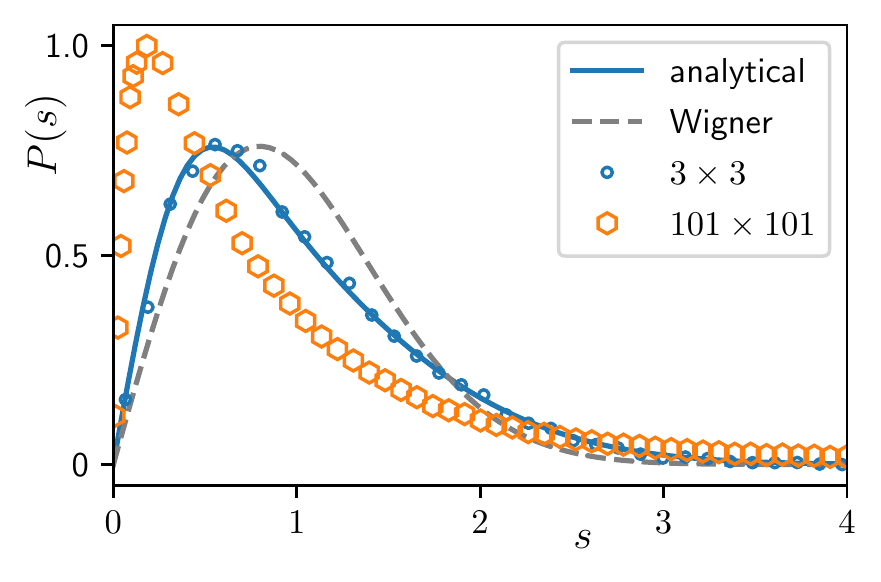}
	\caption{\revisedd{Numerically computed} normalized spacing 
	distribution $P(s)$ \revisedd{(represented with blue open circles)} 
	for $ 3\times 3 $ 
		palindromic symmetric Toeplitz matrices is plotted along with the 
		analytical 
		result (\ref{eq:spac_pst_avspac1})  \revisedd{(represented by the 
		solid blue line)}. The Wigner distribution \revisedd{(represented 
		by 	dashed line)} is plotted to show the extent of deviation. 
		\revised{The nearest neighbour spacing distribution for non-zero 
		even-parity eigenvalues for $ 101\times 101$ size matrices are 
		plotted which shows level repulsion \revisedd{(represented by the 
		orange hexagons)}. The 7000 samples of $ 101 \times 101 $ size 
		matrices are utilized to obtain the numerical distribution.}}
	\label{fig:3by3spacing_pst}
\end{figure}

The spacing between the eigenvalues $E_2$ and $E_3$ in terms of matrix 
elements is,
\begin{equation}\label{eq:spacing_derived}
	s = (E_3-E_2) =\sqrt{a^2+8b^2}.
\end{equation}
Using substitution $ a= r\cos \theta $ and $ \sqrt{8} b=r \sin \theta $, 
the spacing distribution therefore can be calculated as,
\begin{equation}\label{eq:spacing_dist_explicit_pst}
	\begin{aligned}
		P(s) &=\frac{2\sqrt{5} A}{\pi} \int \exp[-A(5r^2-\frac{9}{2}r^2 
		\sin^2\theta)] \delta(s-r) \frac{r}{\sqrt{8}} \ud a \ud b\\
		&= \sqrt{10} A \ s \  I_0(\frac{9}{4}As^2) e^{-\frac{11}{4}As^2}.  
	\end{aligned}
\end{equation}
In rescaled variable $ x = s/\bar{s} $ the distribution 
obtained in (\ref{eq:spacing_dist_explicit_pst}) 
transforms to, 
\begin{equation}\label{eq:spac_pst_avspac1}
	P(x) = \frac{2 \sqrt{10}  E\left(\frac{9}{10}\right)^2 
	}{\pi} x e^{-\frac{11  E\left(\frac{9}{10}\right)^2}{2 \pi }x^2} 
I_0\left(\frac{9 x^2 E\left(\frac{9}{10}\right)^2}{2 \pi }\right)
\end{equation}
where $ E\left(\frac{9}{10}\right) $ is the complete Elliptic integral 
defined as, $ \int_0^{\pi/2 } \sqrt{1-m \sin ^2 \theta} \, 
d\theta $ and $ I_0(z) $ is the modified Bessel function of the first kind 
of order zero.   The spacing distribution is plotted in Fig. 
(\ref{fig:3by3spacing_pst}) and compared with the numerical distribution.
We can clearly see the level repulsion in this ensemble but the form of 
which is different from the Wigner distribution. The extent of deviation 
from this universal result of Wigner is quite pronounced in Fig. 
(\ref{fig:3by3spacing_pst}). 

\revised{We further study the spacing distribution for large $ n 
$-dimensional 
matrices of this class numerically. Like symmetric circulant case, the 
palindromic symmetric Toeplitz matrices commutes with a parity like 
operator defined as,
\[ \sigma = \begin{bmatrix}
0 & 0 & 1\\
0 & 1 & 0\\
1 & 0 & 0
\end{bmatrix} \text{ with } \sigma^2 = \begin{bmatrix}
1 & 0 & 0\\
0 & 1 & 0\\
0 & 0 & 1
\end{bmatrix} \]
This implies that spectrum is divided into two parts, one with even 
parity 
and the other with odd parity. For all the odd-dimension matrix of this 
class, one eigenvalue will always be zero. Therefore, we study the 
nearest neighbour distribution of non-zero even parity eigenvalues. The 
numerically obtained probability distribution function with average 
spacing of one is plotted in the Fig. \ref{fig:3by3spacing_pst}. This 
distribution though shows level repulsion, is very different from Wigner 
distribution. The importance of symmetry reduced spectrum to study the 
spacings can not be overemphasized as one would obtain the Poisson 
distribution in place of level repulsion due to mixing of two parity 
sectors of the spectrum.

}

\section{Summary and discussion}
In this paper, we have studied patterned random matrices which are real 
symmetric with only $ (n+1)/2 $ independent entries in contrast to $ 
n(n+1)/2 $ independent entries allowed for real symmetric matrices. We 
mainly focussed on the calculation of spacing distribution and 
restricted our attention to $ 3\times 3 $ matrices to derive all the 
distributions analytically. We have analytically shown and numerically 
verified that spacing distributions come out to be very different 
ranging from a heavy tailed one for reverse circulant matrices with 
additional zeros to level repulsion coming from the additional 
structural constraint put on these matrices. The $ n\times n $ 
generalizations of symmetric circulant is fairly simple and spacing 
becomes Poisson distributed, but the similar level of analytical 
understanding is beyond the reach of methods used in this work. The 
biggest bottleneck presents itself in terms of absence of eigenvalue 
formulae in terms of matrix elements for PST ensembles. 
\revised{ Therefore, we studied the large dimensional generalizations 
numerically 
and have shown that in large $ n $-limit, constrained reverse circulant 
matrices and symmetric circulant matrices approach to the same spacing 
distribution which in this case is Poisson and not Wigner as expected 
for 
Gaussian orthogonal ensemble. The palindromic symmetric toeplitz 
matrices 
on the other hand despite having exactly same number of elements as the 
previous two cases display level repulsion. The distribution itself is 
different from Wigner distribution. To summarize, the patterned random 
matrices continue to surprise in terms of the behaviour of spacing 
distribution away from universality despite having the same symmetry 
class and equal number of independent elements.   }
\section*{References}

\bibliographystyle{iopart-num-mod}
\bibliography{abbr,extracted,references}

\providecommand{\newblock}{}
\begin{thebibliography}{10}
\providecommand{\eprint}[2][]{\url{#2}}

\bibitem{Wig1957a}
Wigner E~P 1957 {Statistical properties of real symmetric matrices with many
  dimensions} {\em Can. Math. Congr. Proc.\/}  174

\bibitem{GuhMueWei1998}
Guhr T, M\"uller-Groeling A and Weidenm\"uller H~A 1998 Random-matrix theories
  in quantum physics: common concepts {\em Phys.~Rep.\/} {\bf 299} 189--425

\bibitem{RMTScholarpedia}
Fyodorov Y 2011 {R}andom matrix theory {\em Scholarpedia\/} {\bf 6} 9886

\bibitem{BohGiaSch1984}
Bohigas O, Giannoni M~J and Schmit C 1984 Characterization of chaotic quantum
  spectra and universality of level fluctuation laws {\em Phys.~Rev.~Lett.\/}
  {\bf 52} 1--4

\bibitem{BerTab1977}
Berry M~V and Tabor M 1977 Level clustering in the regular spectrum {\em
  Proc.~R.~Soc.~Lon.~A\/} {\bf 356} 375--394

\bibitem{Dys1962b}
Dyson F~J 1962 The threefold way. {Algebraic} structure of symmetry groups and
  ensembles in quantum mechanics {\em J.~Math.~Phys.\/} {\bf 3} 1199--1215
  0022-2488

\bibitem{AltZir1997}
Altland A and Zirnbauer M~R 1997 Nonstandard symmetry classes in mesoscopic
  normal-superconducting hybrid structures {\em Phys.~Rev.~B\/} {\bf 55}
  1142--1161

\bibitem{Bos2018}
Bose A 2018 {\em Patterned random matrices\/} (Chapman and Hall/CRC)

\bibitem{BosSahSen2021}
Bose A, Saha K and Sen P 2021 Some patterned matrices with independent entries
  {\em Random Matrices: Theory and Applications\/} {\bf 10} 2150030

\bibitem{BosSahSen2022}
Bose A, Saha K and Sen P 2022 Erratum: Some patterned matrices with independent
  entries {\em Random Matrices: Theory and Applications\/} {\bf 11} 2292001

\bibitem{AdhSah2017}
Adhikari K and Saha K 2017 Fluctuations of eigenvalues of patterned random
  matrices {\em Journal of Mathematical Physics\/} {\bf 58} 063301

\bibitem{BlaBorBos2021}
Blackwell K, Borade N, Bose A, VI C~D, Luntzlara N, Ma R, Miller S~J, Mukherjee
  S~S, Wang M and Xu W 2021 Distribution of eigenvalues of matrix ensembles
  arising from wigner and palindromic toeplitz blocks

\bibitem{JaiSri2008}
Jain S~R and Srivastava S~C~L 2008 {Random cyclic matrices} {\em Phys. Rev.
  E\/} {\bf 78} 036213

\bibitem{JaiSri2009}
Jain S~R and Srivastava S~C~L 2009 {Random matrix theory for pseudo-Hermitian
  systems: Cyclic blocks} {\em Pramana-J. Phys.\/} {\bf 73} 989

\bibitem{SriJai2012}
Srivastava S~C~L and Jain S~R 2012 Random reverse-cyclic matrices and screened
  harmonic oscillator {\em Phys. Rev. E\/} {\bf 85}(4) 041143

\bibitem{ShuSad2015}
Shukla P and Sadhukhan S 2015 Random matrix ensembles with column/row
  constraints: I {\em J.~Phys.~A - Math.~Theor.\/} {\bf 48} 415002

\bibitem{SadShu2015}
Sadhukhan S and Shukla P 2015 Random matrix ensembles with column/row
  constraints: {II} {\em J.~Phys.~A - Math.~Theor.\/} {\bf 48} 415003

\bibitem{GluEisFar2019}
Gluza M, Eisert J and Farrelly T 2019 {Equilibration towards generalized Gibbs
  ensembles in non-interacting theories} {\em SciPost Phys.\/} {\bf 7}(3) 38

\bibitem{ManSriJai2011}
Manikandan K, Srivastava S~C~L and Jain S~R 2011 {Biased random walks on a
  disordered one-dimensional lattice} {\em Phys. Lett. A\/} {\bf 375} 368

\bibitem{Dem2017}
Demidenko E 2017 {Applications of Symmetric Circulant Matrices to Isotropic
  Markov Chain Models and Electrical Impedance Tomography} {\em Adv. in Pure
  Math.\/} {\bf 7} 188

\bibitem{KarSchUeb2003}
Karner H, Schneid J and Ueberhuber C~W 2003 {Spectral decomposition of real
  circulant matrices} {\em Linear Algebra Appl.\/} {\bf 367} 301

\bibitem{MolSok1989}
Molinari L and Sokolov V~V 1989 Level repulsion for band 3 $times$ 3 random
  matrices {\em Journal of Physics A: Mathematical and General\/} {\bf 22} L999

\bibitem{BerShu2009}
Berry M~V and Shukla P 2009 Spacing distributions for real symmetric 2 $ times
  $ 2 generalized gaussian ensembles {\em J.~Phys.~A - Math.~Theor.\/} {\bf 42}
  485102

\bibitem{BosChaGan2003}
Bose A, Chatterjee S and Gangopadhyay S {Limiting spectral distribution of
  large dimensional random matrices} {\em J. Indian Statist. Assoc.\/} {\bf 41}
  221 -- 259

\bibitem{BryDemTie2006}
Bryc W, Dembo A and Jiang T 2006 {Spectral measure of large random Hankel,
  Markov and Toeplitz matrices} {\em The Annals of Probability\/} {\bf 34} 1 --
  38

\bibitem{MasMilSin2007}
Massey A, Miller S~J and Sinsheimer J 2007 Distribution of eigenvalues of real
  symmetric palindromic toeplitz matrices and circulant matrices {\em Journal
  of Theoretical Probability\/} {\bf 20} 637--662 1572-9230

\end{thebibliography}

\end{document}